**Variational Learning of Physical Intuition from a Few Observations: Charting Manifolds of Variational Physics**

Jingruo Peng[1], Shuze Zhu*[1]

[1]Center for X-Mechanics, Institute of Applied Mechanics, Zhejiang University, Hangzhou 310000, China

*To whom correspondence should be addressed. E-mail: shuzezhu@zju.edu.cn

## Abstract

Humans often generalize physical outcomes from few observations, a desirable capacity known as physical intuition. We show that it can be computationally approached through charting the manifolds of variational physics. We conceive a variational learning framework, where small neural networks trained from merely two or three instances delineate an observation-spanned solution manifold, enabling generalization in unseen scenarios. As demonstrated across classical and quantum regimes including strongly correlated molecules, generalization emerges far beyond the training conditions. This generalization is explained by a unified theory: the Euler–Lagrange operator must be stationary with respect to the observation parameters along the solution manifold, whose complexity necessitates a critical network capacity below which generalization fails. Our framework establishes a principled route to solving families of variational problems from a few instances, and shows how physical intuition can be approached through harnessing variational physics.

*Introduction* − Many fundamental equations in physics—from classical mechanics to quantum physics—are consequences of the variational principle. Optimal physical state minimizes a functional. The arc of a projectile, the deflection of an elastic plate, and the ground-state wavefunction of a molecule are all such solutions. However, a new set of initial or boundary conditions typically demands solving the problem anew. It is remarkable that humans and other biological agents navigate such challenges with apparent ease. Ancient hunters, after only a few attempts, develop an accurate sense of where a spear lands, and humans form such intuitions from a few observations [1–3]. This capacity is known as physical intuition.

Computationally realizing such efficient generalization remains a challenge. While data-driven models have achieved remarkable success across many domains [4–16], their reliance on massive datasets questions whether scale alone can substitute for the efficiency of physical reasoning [17,18], motivating principles that enable rapid generalization from extremely limited observations [19–21].

Here, we show that when the governing variational physics is elevated into a higher-dimensional manifold, this capacity for generalization admits a concrete computational realization—an efficient one: a neural network trained from merely two to three instances learns to predict optimal states for unseen conditions by charting the local geometry of the solution manifold. In essence, the trained network solves a family of variational problems, akin to an observer who, by learning a few cases, gleans how solutions vary under changing circumstances.

The key concept is that the solution manifolds of variational physics can be spanned by many choices of observation features—any observable can parameterize a manifold of optimal states, none a priori privileged—and exploring along a connected path on such a manifold enables generalization from limited training conditions. We realize this through a variational learning framework that charts the manifold projected onto particular observation features. We adopt a "train–freeze–predict" strategy: networks are trained on two or three observations, frozen, and evaluated on unseen cases. This unsupervised instance-efficient approach differs fundamentally from conventional physics-guided machine learning [22–26]. Unlike physics-informed neural networks (PINNs) that solve a single instance, our method learns the variational geometry of the solution manifold in higher dimensions, enabling generalization across instances. We find that robust generalization can emerge from as few as two highly similar observations, and that generalization improves dramatically with the number of informative instances. We also identify a critical network size (approximately 100–200 parameters for simpler systems, rising to 1000–3000 for the strongly correlated $N_2$ molecule) below which this capability fails. These findings are explained by a unified theory derived from the stationarity of the Euler–Lagrange equation along the manifold. Our work thus proposes a principled framework for solving families of variational problems, and offers a mechanistic perspective on how physical understanding might emerge from sparse experience.

*Principle of variational learning* — The optimal state $y(x)$ of many physical systems minimizes a functional $F[y]$, where $x$ denotes generalized spatial-temporal coordinates. Embedding the observation parameter $\zeta$, which specifies the observation (e.g., boundary positions, load magnitudes, molecular geometries), elevates the solution $y(x,\zeta)$ to a $\zeta$-dependent higher-dimensional manifold spanned by $\zeta$. At the optimum, the first variation vanishes for any perturbation $\delta y$, controlled by an infinitesimal scalar $\tau$:

$$\lim_{\tau\to 0}\frac{1}{\tau}(F[y+\tau\delta y]-F[y])=0. \tag{1}$$

Single-instance solution comes from a fixed $\zeta_i$, where one can approximate the optimal state with a neural network $y=NN(x,\zeta;W)$ by minimizing the functional $L_i=F[NN(x,\zeta_i;W)]$ with respect to the parameters $W$. Without $\zeta$ embedding, single-instance approach corresponds to conventional settings in physics-guided machine learning.

The core of *intuition manifold learning*, however, is to seek a single fixed set of parameters $W^*$ such that the network $NN(x,\zeta;W^*)$ approximates the optimal state for a range of $\zeta$. The family of optimal states forms a solution manifold spanned by $\zeta$, and training on a few instances must constrain the shared parameterization to delineate this manifold. For the trained network, the functional $F$ should ideally be stationary for any admissible $\zeta$ on the solution manifold. Mathematically, we require for all $\zeta_k$ on the solution manifold:

$$\sum_k\left|\lim_{\tau\to 0}\frac{1}{\tau}(F[NN(x,\zeta_k;W)+\tau\delta NN]-F[NN(x,\zeta_k;W)])\right|\cong 0. \tag{2}$$

Enforcing this condition directly across all $\zeta$ is computationally intractable. For two physically similar scenarios $\zeta_k$ and $\zeta_j$, however, the optimal values of $F$ should be close. This motivates a proximal condition: the difference between the functionals evaluated at nearby observations should vanish to first order. The corresponding finite-difference approximation is

$$\sum_{k,j} \left| \lim_{\tau_{kj}\to 0} \frac{1}{\tau_{kj}} \left( F\left[NN\left(x, \zeta_j; W\right)\right] - F[NN(x, \zeta_k; W)] \right) \right| \cong 0, \tag{3}$$

where $\tau_{kj}$ quantifies the small difference between $\zeta_k$ and $\zeta_j$. Equation (3) reveals what the training protocol must achieve: in addition to minimal variational loss at each training condition, the loss must also change minimally between nearby conditions. This requires the network to learn how the optimal state varies with $\zeta$, thereby charting the correct form of manifold spanned by $\zeta$.

*Variational learning protocol* $-$ We implement Eq. (3) via an alternating optimization protocol: the training minimizes $L_k = F[NN(x, \zeta_k; W)]$ on odd steps and $L_j = F[NN(x, \zeta_j; W)]$ on even steps, while continuously updating the shared parameters $W$ across steps. This process converges to a single parameter set $W$ that not only simultaneously makes $F$ stationary for both $\zeta_k$ and $\zeta_j$, but also encodes how the optimal physical state should vary with $\zeta$. The protocol extends to more observations by cycling through them. Alternating steps are crucial: parameters sensitive to $\zeta$ receive larger updates, thereby extracting the local direction of the corresponding solution manifold. The few training instances serve as anchor points on manifold—mathematically analogous to boundary conditions—constraining the shared parameterization to a specific local branch and enabling prediction for unseen $\zeta$ without further optimization.

*Performance metric* $-$ After training, we freeze $W$ and evaluate predictions on unseen observations $\zeta_p$. Coefficient of determination ($R^2$) is used for field outputs and relative error for scalar outputs. Robust physical intuition corresponds to $R^2 > 0.9$ or relative error < 10% unless otherwise noted.

*Strongly Correlated Quantum Physics* $-$ We first demonstrate on a challenging quantum many-body system: the triple-bond nitrogen molecule ($N_2$). With 14 electrons, $N_2$ is strongly correlated, and standard quantum chemistry methods fail at stretched geometries [27]. This makes $N_2$ an ideal testbed. One of its solution manifolds—the family of ground state wavefunctions parameterized by bond length R—spans qualitatively distinct electronic regimes, from the triple-bonded stable state to dissociated state where the molecule disintegrates. These regimes correspond to distinct local solution manifolds.

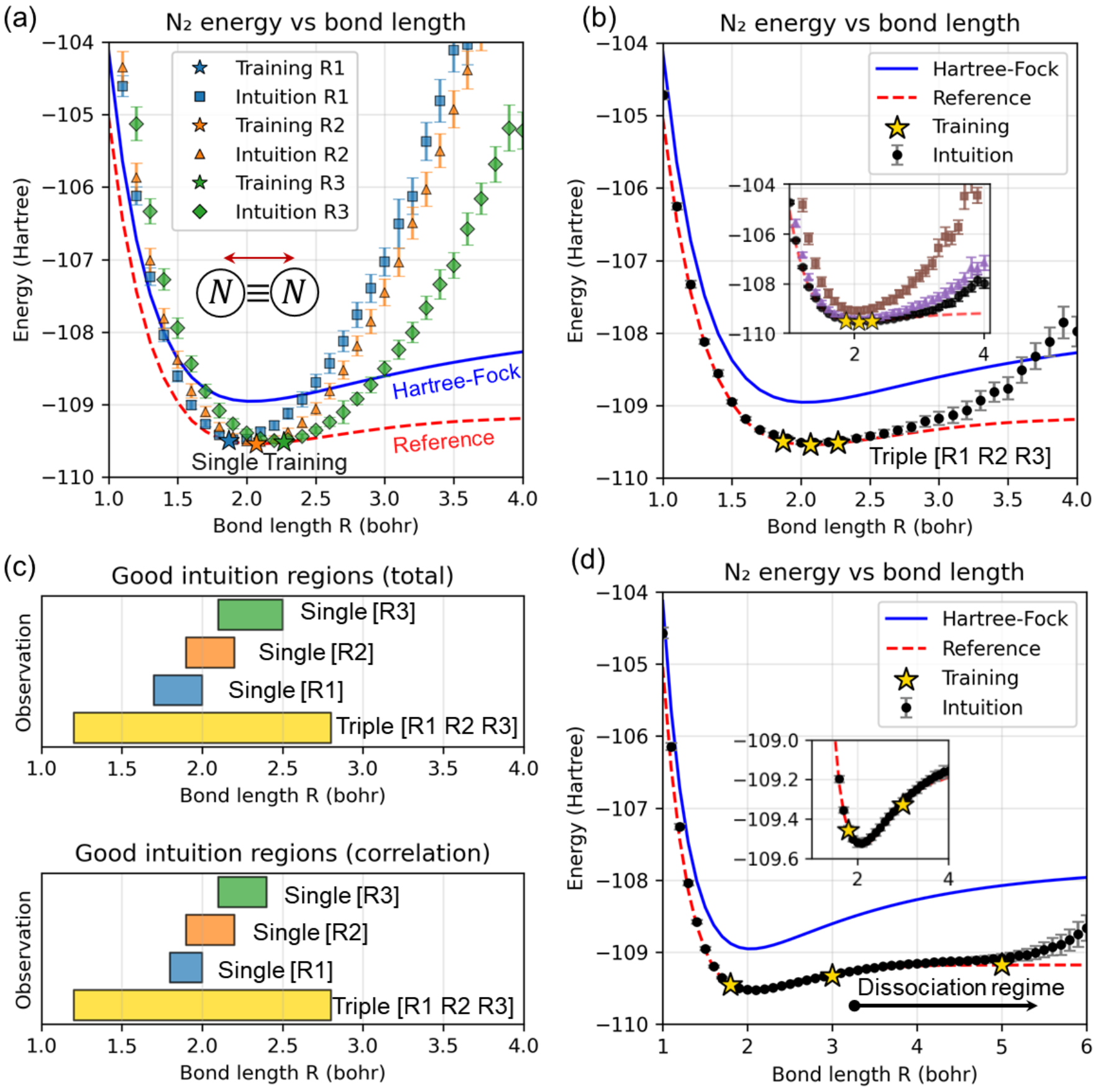

FIG. 1. Variational learning of $N_2$ potential energy. (a) Single-observation models: predictions (markers with error bars) at training bond lengths R = 1.868, 2.068, 2.268 bohr (stars) compared with Hartree-Fock (solid blue) and high-accuracy reference (dashed red). Each single model only generalizes locally. (b) Triple-observation model trained on the same three bond lengths (alternating protocol) reproduces the reference curve over a wide range with small error bars. The inset shows that finer grasp on the training observations leads to stronger intuitions. (c) Good intuition regions for total energy (relative error < 0.1%) and for correlation energy (relative error < 15%). Bars correspond to single (short) and triple (long) training. (d) Using manifold-connecting technique, triple sparse observations (R = 1.8, 3.0, 5.0 bohr) reconstruct the entire energy curve, including the dissociated regime. The inset shows that the network correctly predicts the energy minimum even though no training point lies exactly at equilibrium.

We employ FermiNet [28] as neural network ansatz for many-electron wavefunction. In standard use, FermiNet is optimized by variational Monte Carlo (VMC) for a single geometry. We modify the training logic with the alternating protocol, thereby imposing an additional variational constraint across geometries on top of the standard VMC optimization. The loss function at each training step is the energy expectation value for each geometry. After training, we freeze the network parameters and evaluate ground-state energy using Monte Carlo sampling on a dense grid of bond lengths.

Figure 1(a–b) displays the resulting energy curves together with high-accuracy reference and the Hartree-Fock energy [28,29]; their difference is the correlation energy. Here we alternate among three similar bond lengths R = 1.868, 2.068, 2.268 bohr (close to equilibrium) for the triple-observation model, and also train three single-observation

baselines on each bond length individually. Each single-observation model [Fig. 1a] achieves high accuracy only in a narrow interval around its training point, with rapidly increasing uncertainty outside. In contrast, the triple-observation model [Fig. 1b] reproduces the reference curve accurately over a wide range with small error bars, reflecting that the shared representation stabilizes the prediction far from the training points. In the ideal limit where the learned solution coincides with the exact solution, predictions can be accurate everywhere; approximating this ideal scenario, finer grasp on the training points from a finite-capacity model leads to stronger intuitions [Fig. 1b, inset]. Figure 1c quantifies the good-intuition intervals: the triple-observation model attains a continuous interval of 1.4 bohr under both total-energy (error $< 0.1\%$) and correlation-energy (error $< 15\%$) criteria, whereas each single-observation model covers only about 0.3 bohr. We have verified the robustness to batch size and training-point placement: Training from compressed side or from the stretched side still captures the equilibrium minimum, confirming that the learned representation is not merely interpolating.

To test generalization across qualitatively different regimes, we trained another triple-observation model on bond lengths $R = 1.8$, 3.0, and 5.0 bohr, which span the bonded and dissociated regions. The resulting network reconstructs the entire energy curve from equilibrium to the dissociated limit with high fidelity [Fig. 1d], from three sparse observations. The network correctly predicts the energy minimum even though no training point lies exactly at equilibrium ($R \approx 2.068$ bohr). When the two outer points are used without the intermediate one ($R = 1.8$ and 5.0 alone), the network fails, revealing that these distinct local manifolds, separated by triple-bond dissociation, require an intermediate anchor point to bridge. The intermediate point connects the local manifolds, indicating how the framework can detect physical regime transitions through generalization failure.

Additional experiments show that the alternating protocol consistently outperforms a joint-loss baseline, especially when the training observations are closely spaced, confirming that the protocol resolves manifold geometry that a fixed joint loss cannot. We also identify a critical network capacity (∼1000–3000 parameters for FermiNet) below which generalization fails, reflecting the complexity of the many-electron wavefunction manifold. Together, these results establish variational learning as an instance-efficient route to master quantum variational problems even for strongly correlated systems. Simpler quantum systems ($H_2$ and quantum harmonic oscillators) that serve as learning dynamics benchmarks are analyzed in Appendices A and B.

*Classical Physics* — We next apply the framework to a challenging classical variational problem with no closed-form analytical solution. A thin square plate of side length $b$, clamped on all four edges, is subjected to a uniformly distributed pressure $p$ [Fig. 2a, top]. The plate undergoes large bending deflections, requiring geometrically nonlinear von Kármán plate theory [30]. The equilibrium state minimizes the total potential energy functional. The learned output consists of three displacement fields $U(x,y)$, $V(x,y)$, $W(x,y)$, a high-dimensional function-space target parameterized by two observation features $(p,\ b)$.

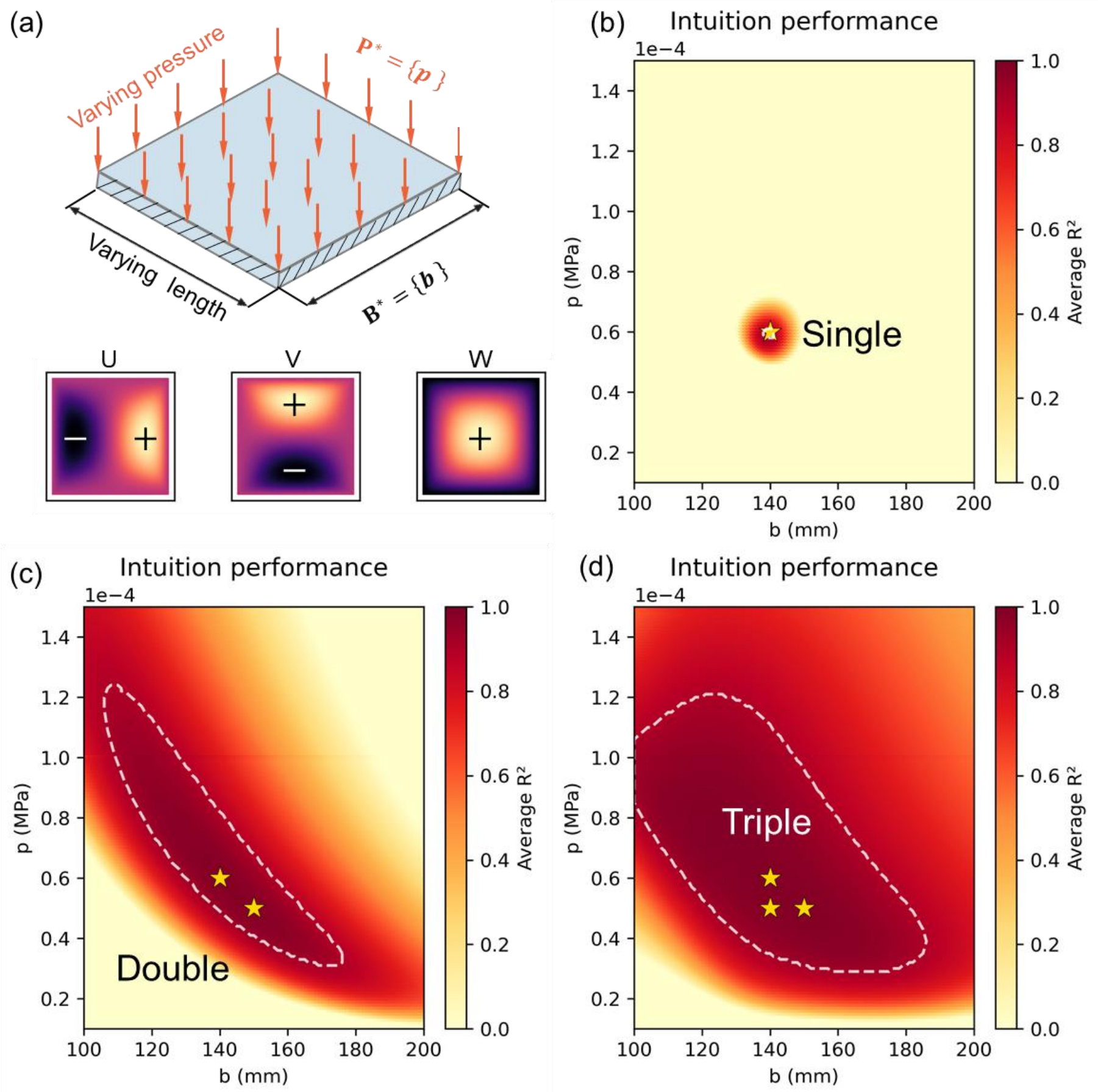

FIG. 2. Variational learning of a nonlinear elastic plate. (a) Top: Schematic of a thin square plate (side length $b$) clamped on all edges and subjected to uniform pressure $p$. Bottom: Finite-element solutions for a representative load-geometry pair, illustrating distinct symmetries and sign patterns in the in-plane displacements $(U,V)$ and the out-of-plane deflection $(W)$. (b-d) Good-intuition region ($R^2$>0.9, enclosed by dashed white) from single-observation (b), double-observation (c), and triple-observation (d) training, where the training observations are marked by stars.

We train three independent neural networks, one per displacement component, using the alternating protocol. The loss is the total potential energy evaluated by numerical integration over random collocation points. After training, we freeze the parameters and evaluate generalization on a dense grid of $(p,b)$ values. A point is considered within good-intuition region when the average $R^2$ over the three displacement components exceeds 0.9. Finite-element solutions for a representative case [Fig. 2a, bottom] illustrate the distinct symmetries and sign patterns of the displacement fields.

Figure 2(b-d) shows the results. A model trained on a single observation $(p, b) = (0.6 \times 10^{-4}, 140)$ yields an extremely localized region of accuracy: only 14 grid points near training point satisfy the $R^2$>0.9 criterion [Fig. 2b]. Adding a second observation $(0.5 \times 10^{-4}, 150)$ [Fig. 2c] expands this region to 1968 grid points, with the long axis aligning along the line connecting the two training points, a clear signature that the network has learned the local direction of the solution manifold. Adding a third observation $(0.5 \times 10^{-4}, 140)$ [Fig. 2d] further expands the region to 4618 grid points. These results show that the variational protocol distinguishes similar displacement fields of these training points, resolving their high-dimensional solution manifold. Additional inspection shows that at the $R^2$>0.9 boundary, the triple-observation model yields qualitatively correct fields, the single fails, and the double is qualitatively correct but inaccurate.

We also find that variational training consistently outperforms joint-loss baseline, especially for closely spaced training points. The critical total parameter count for the triple-observation training is around 100–150: networks with 39 or 93 parameters produce no good-intuition region, while those with 165 or more succeed; below this threshold, the predicted fields lose the correct symmetry and magnitude. Two additional classical problems (brachistochrone and projectile motion) confirm the same picture (Appendices C and D). Together, variational learning can solve a family of variational physics problems across diverse physical regimes, generalizing physical intuition.

*Unified Generalization Theory* – All optimal states in these systems minimize an action functional:

$$S[y; \zeta] = \int dx\, F(x, y(x; \zeta), y'(x; \zeta); \zeta), \tag{4}$$

where $y(x; \zeta)$ is the $\zeta$-embedded physical state, and the specific form of $F$ encodes quantum or classical physics. For example, in quantum mechanics (molecule & oscillator), the ground state minimizes the energy expectation $\langle\psi|H|\psi\rangle$ under normalization. This is equivalent to extremizing $S[\psi] = \int dx \left[\frac{\hbar^2}{2m}\psi^{*\prime} \cdot \psi' + \psi^* V\,\psi - \lambda\psi^*\psi\right]$, whose Euler-Lagrange equation yields the Schrödinger's equation $H\psi = \lambda\psi$.

The stationary condition is the Euler-Lagrange (EL) equation:

$$\frac{\delta S}{\delta y} \equiv \frac{\partial F}{\partial y} - \frac{d}{dx}\frac{\partial F}{\partial y'} = 0 \tag{5}$$

Consider a base observation $\zeta_0$ and a perturbed one $\zeta = \zeta_0 + \delta\zeta$. Taylor expansion gives $y(x;\zeta) \cong y(x;\zeta_0) + \frac{\partial y}{\partial \zeta}\,\delta\zeta$. Enforcing Eq. (5) at $\zeta$ and using optimality at $\zeta_0$ (i.e., $\left.\frac{\delta S}{\delta y}\right|_{\zeta_0} = 0$) yields:

$$\frac{\partial}{\partial \zeta}\left(\frac{\delta S}{\delta y}\right) \equiv \frac{\partial}{\partial \zeta}\left(\frac{\partial F}{\partial y} - \frac{d}{dx}\frac{\partial F}{\partial y'}\right) = 0 \tag{6}$$

Equation (6) is the fundamental condition enabling generalization: the Euler-Lagrange operator must be stationary with respect to observation parameter along the solution manifold, so that a solution optimal at a few $\zeta_k$ remains optimal at unseen $\zeta_p$. Our alternating training protocol [Eq. (3)] is a finite-difference approximation to this condition. Minimizing $L_k$ and $L_j$ alternately with shared parameters drives $\frac{\delta S}{\delta y} = 0$ at both $\zeta_k$ and $\zeta_j$, while simultaneously minimizing their difference so that $\frac{\partial}{\partial \zeta}\left(\frac{\delta S}{\delta y}\right) \approx \frac{\left(\frac{\delta S}{\delta y}\right)|_{\zeta_j} - \left(\frac{\delta S}{\delta y}\right)|_{\zeta_k}}{\zeta_j - \zeta_k} = 0$. Alternating loss forces the network to repeatedly adapt to each $\zeta$, implicitly penalizing large changes in the operator across $\zeta$. In contrast, joint-loss (e.g., $L_k + L_j$) baseline averages the gradients and dilutes the directional derivative across $\zeta$. The reasoning extends to multiple variables or systems of equations.

*Threshold for Robust Generalization* — Strong generalization requires the network to express the $\zeta$-projected solution manifold, satisfying both $\frac{\delta S}{\delta y} = 0$ and $\frac{\partial}{\partial \zeta}\left(\frac{\delta S}{\delta y}\right) = 0$. Below a critical network capacity, the network cannot meet these requirements, and thus fails to generalize.

We test this across three quantum and three classical systems. For each, we train models on three observations using the alternating protocol, and measure the generalization performance as a function of trainable parameters. Figure 3 plots this relationship. In all cases, generalization performance remains low until a critical size, then rises sharply and saturates. Below this threshold, models are locally plausible but lack correct $\zeta$-projected variation. Relatively simpler systems show a sharp transition near 100-200 parameters [Fig. 3b-3f]. For strongly correlated $N_2$ molecule, which learns a far more complex wavefunction manifold, the transition shifts to approximately ~1000-3000 parameters [Fig. 3a]. The consistent appearance of a threshold across physical regimes and computational implementations validates our theory: the critical size reflects the intrinsic complexity of the solution manifold.

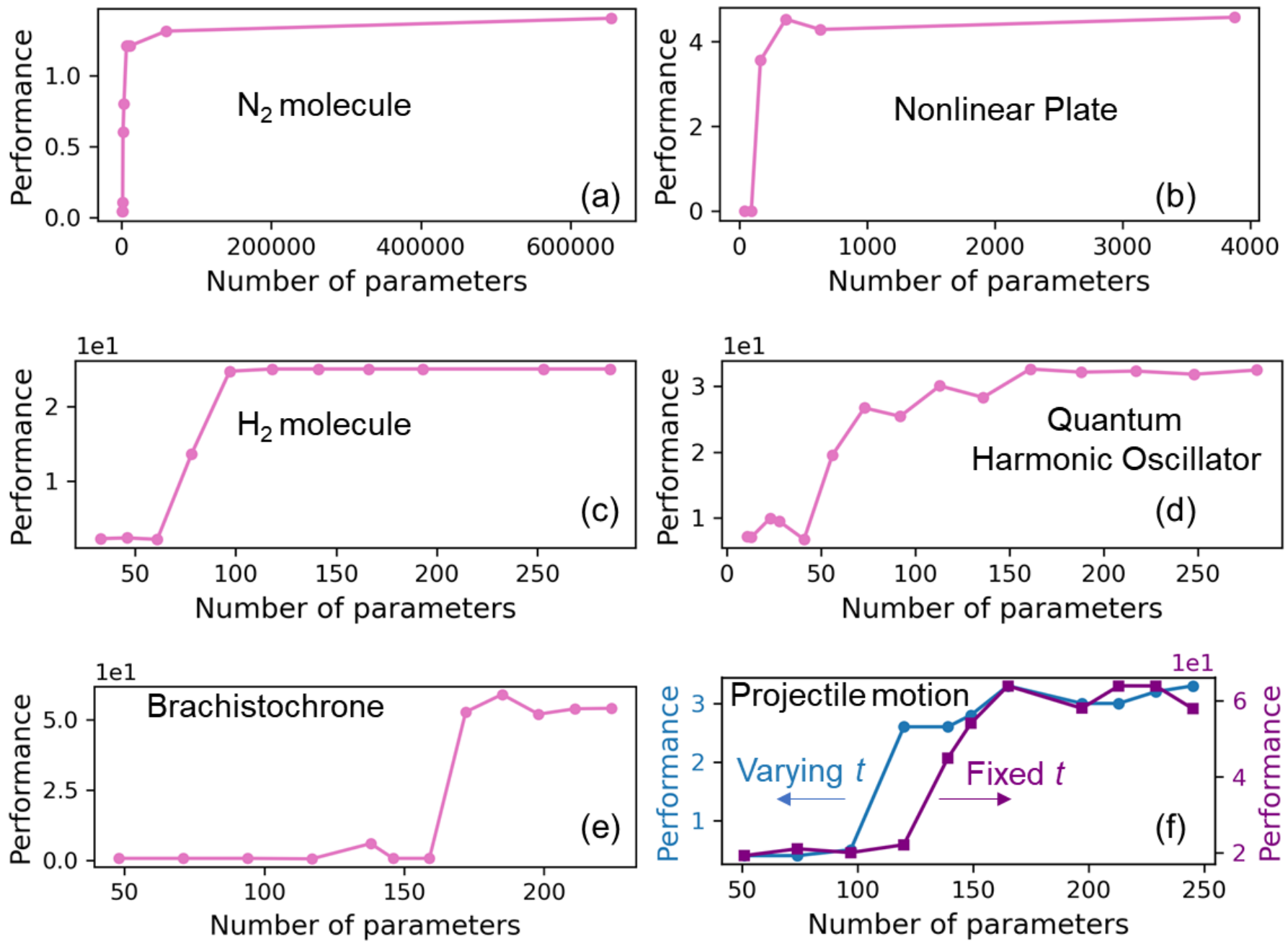

FIG. 3. Critical network size for robust generalization. Generalization performance vs. trainable parameters for six systems, all trained via variational learning on three observations. (a) $N_2$. (b) Nonlinear elastic plate. (c) $H_2$. (d) Quantum harmonic oscillator. (e) Brachistochrone. (f) Projectile motion (showing two curves: fixed flight time (squares) and varying flight time (circles)). In all cases, generalization remains low below a critical size, then rises sharply and saturates. Threshold: ~100-200 parameters for simpler problems; ~1000-3000 for $N_2$, reflecting manifold's complexity.

*Discussion* — The key insight of this work is that variational physics can be approached from a high-dimensional perspective: the solution manifold is an infinite-dimensional geometric object that must be projected to $\zeta$ for practical use, and its connectivity, branching, and regime transitions must be discovered from anchor points. The choices of $\zeta$ are not unique; other observables yield equally valid manifold perspectives, making our approach broadly applicable. This elevation in dimension—simultaneously solving at anchor points and charting the manifold that spans them—is what enables generalization. The variational principle constrains admissible solutions; the alternating protocol, guided by $\partial(\delta S/\delta y)/\partial\zeta = 0$, resolves the local geometry of the $\zeta$-projected manifold by probing the observation gradient between anchor points while driving them toward $\delta S/\delta y = 0$, thereby delineating the correct manifold pathway. Neither the manifold nor the anchor points on it are given a priori; both are determined through the protocol. Our approach learns to predict the solution of a family of variational problems through a single unsupervised model, requiring no paired data and only 2–3 instances to anchor the solution branch. The combination—solving problems in high-dimensional manifolds, unsupervised data efficiency, physical grounding, and instance-generalizing—is, to our knowledge, unique among existing physics-guided machine learning. The high-dimensional perspective, together with the theory for charting the resulting manifold, opens an unexplored route to instance-efficient strong generalization.

Several lines of evidence support this conceptual insight. First, under identical conditions, we find that the alternating protocol consistently outperforms joint-loss training, with the largest advantage when training instances are most similar—the regime where subtle observation-dependent differences are critical for resolving the manifold geometry. Second, the $N_2$ molecule reveals the role of manifold connectivity: its wavefunction belongs to distinct local manifolds at short and long bond lengths—two anchors in different regimes fail, while a third bridges them (Fig. 1d). A globally smooth projection thus conceals distinct local manifolds, bridged only by intermediate anchors. Third, the $\zeta$-projected picture implies operational constraints: anchor points must be correctly placed on the manifold and connected, requiring critical network capacity to express the solution manifold. For simpler systems, resolving O(10) degrees of freedom along $x$ and coupling them via continuity in $\zeta$ yields O(100) parameters [Fig. 3]. Its increase for $N_2$ reflects the higher dimensionality of its many-electron wavefunction. Hence, the critical size measures the problem's own complexity, not a universal constant. Finally, the method scales to deeper architectures: residual connections [31] preserve observation-feature information through multiple layers, enabling robust manifold learning even in deep networks.

Learning effectiveness also depends on the informativeness of training resources. We find that physics-rich regimes (e.g., near the equilibrium bond length where interactions are strong), bridging regimes [e.g., Fig. 1d] both enable broader generalization. As training observations are more accurately placed on the connected manifold, the good-intuition domain expands progressively (Fig. 1b), confirming that fidelity to the exact local solution branch is the driver of generalization. These echo human learning, where informative experiences are more potent.

*Conclusions* – We show that strong generalization in variational physics emerges from elevating the problem into a higher-dimensional solution manifold. This is done by embedding observable $\zeta$ into the physical state. The $\zeta$-projected geometry of this manifold is resolved by a variational learning framework, theorized by the stationarity of the Euler-Lagrange operator $\partial(\delta S/\delta y)/\partial\zeta = 0$. As demonstrated across classical and quantum systems, including the strongly correlated $N_2$ molecule, generalization emerges far beyond the training instances. A critical network capacity is required to express the intrinsic complexity of the solution manifold, below which robust generalization fails. This variational-physics-informed machine learning framework establishes a principled route to solving families of variational physics problems from sparse observations. It offers a concrete computational realization of physical intuition, paralleling how informative experiences sharpen understanding in biological learning. It operates with a clear objective functional, complementing approaches that discover such functionals directly from data [32,33].

## Appendix A: Hydrogen molecule

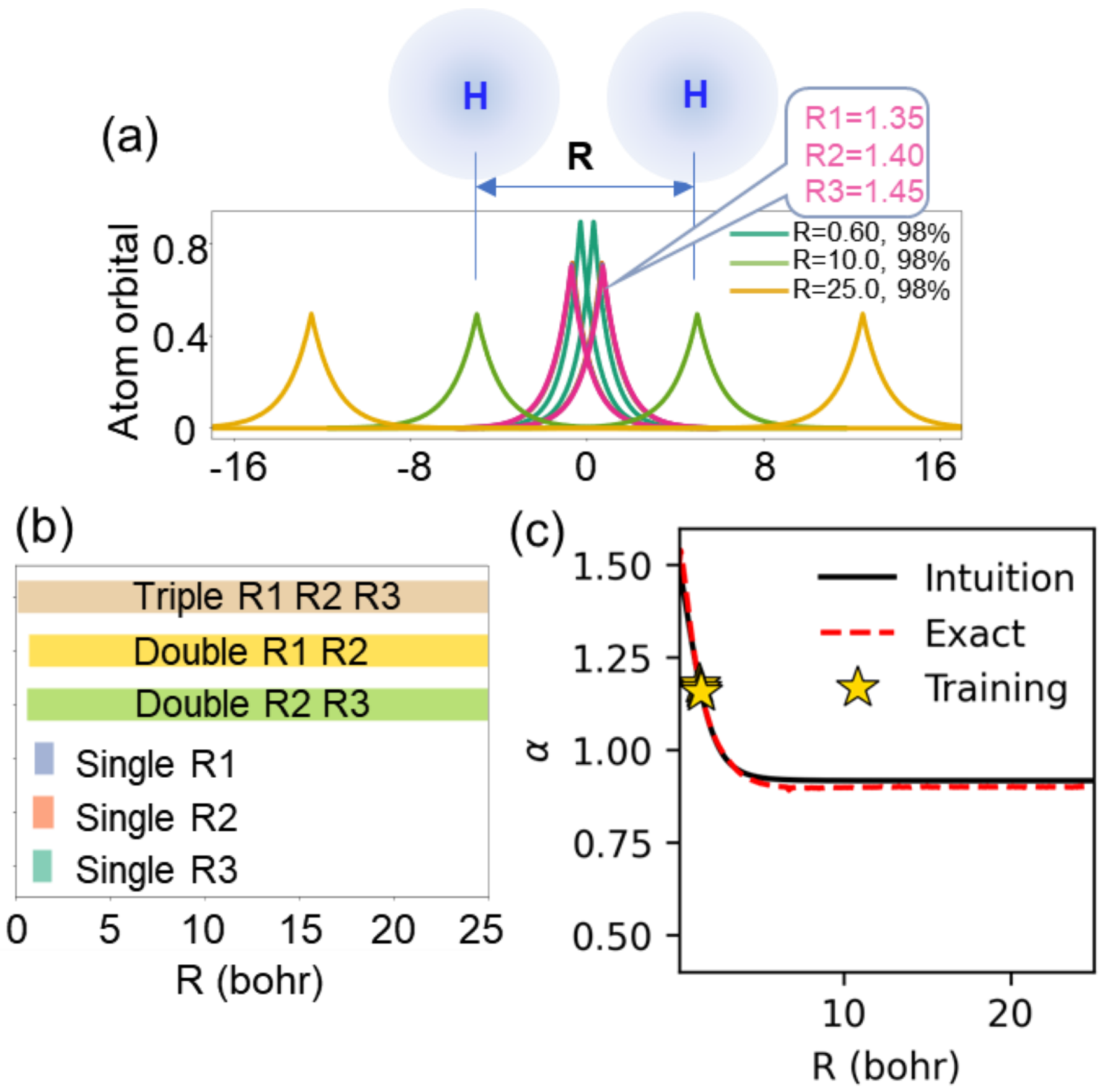

FIG. 4. Accurate physical intuition from a few highly similar observations in quantum molecular systems ($H_2$ molecule). (a) Predicted 1s atomic orbitals at various bond lengths $R$. The network is trained on three closely spaced bond lengths ($R = 1.35, 1.40, 1.45$) and achieves > 98% accuracy even for distant extrapolations (e.g., $R = 0.6, 25$). (b) Range of bond lengths over which the network maintains high accuracy ($R^2$ > 0.9) when trained on one, two, or three observations via variational learning. (c) Predicted screening parameter $\alpha$ as a function of $R$ (black solid line) versus exact solution (red dashed line). Three training points are marked in gold stars.

As a tractable test case, we manually solve a hydrogen molecule using DFT-informed artificial neural network, which learns the screening parameter $\alpha$ in the 1s atom orbital basis function $a(\vec{r}) = \sqrt{\frac{\alpha^3}{\pi}} e^{-\alpha|\vec{r}-\vec{R_a}|}$, as a function of bond length $R$, via an ANN $\alpha = NN(R; W)$. The loss function $L(R; W)$ is the total energy, whose minimization yields the ground state. The loss function can be constructed manually by density functional theory (local density approximation) [34]. The network is trained on only three closely spaced bond lengths ($R = 1.35, 1.40, 1.45$ Bohr), marked in Figure 4(a). Using the variational learning protocol, the network learns to accurately generalize $\alpha(R)$. Remarkably, it achieves ~98% accuracy even for distant extrapolations (e.g., $R = 25$ and $R = 0.6$), as shown by the representative orbitals in Figure 4(a). The prowess of variational learning is quantified in Figure 4(b). We compare the ranges of $R$ over which the network maintains high accuracy ($R^2 > 0.9$) when trained on single, double, or triple observations. The “good-intuition” range expands dramatically as more similar observations are incorporated via variational learning. Finally, Figure 4(c) shows that the full nonlinear dependence $\alpha(R)$ is captured after training on just triple observations. More calculations show that the prowess of variational learning approach appears robust against different activation functions given suitable number of parameters, and variational structure captured from observations in a “physically rich” region (where interactions are strong) is critical for developing stronger intuition.

### Appendix B: Quantum harmonic oscillators

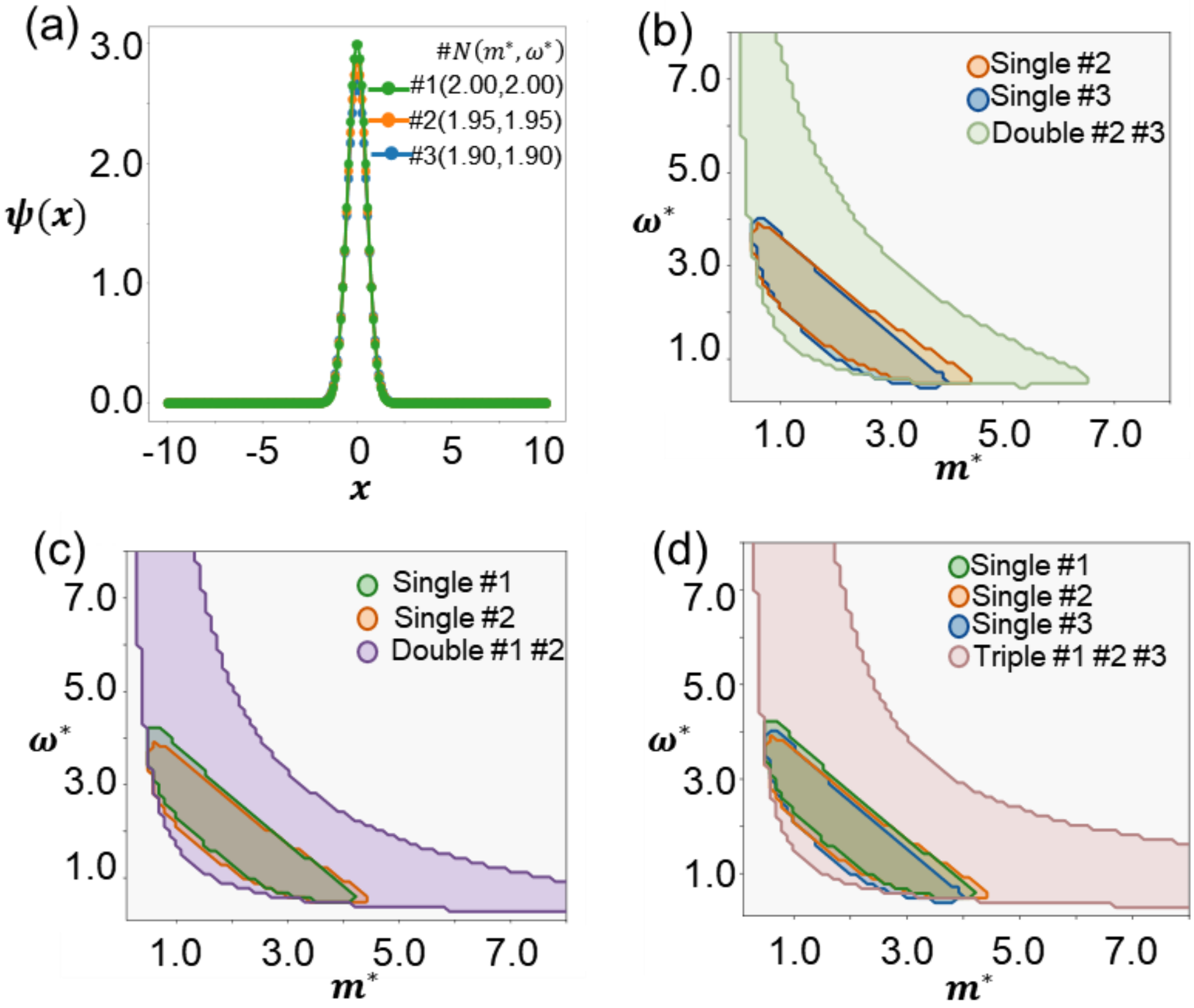

FIG. 5. Variational learning enables extrapolation in a 2D parameter space for quantum harmonic oscillators. (a) Ground-state wavefunctions $\psi(x)$ for the set of three highly similar observations $(m^*, \omega^*)$. (b-d) Regions in $(m^*, \omega^*)$-space where the network maintains high-fidelity predictions ($R^2 > 0.9$) after training. (b) Generalization from single observations (#2, #3) versus variational learning on both (#2, #3). (c) Analogous results for observations #1 and #2. (d) Generalization from each single observation versus variational learning on all three. The intuition region expands substantially with each added observation.

The variational learning can confer robust physical intuition in quantum systems with multiple, continuously variable parameters. A simple illustration of such is quantum harmonic oscillators. We consider the ground state of a one-dimensional quantum harmonic oscillator, $V(x) = \frac{1}{2}m^*(\omega^*)^2x^2$ . The network learns the ground-state wavefunction $\psi(x)$ for any given parameter set $(m^*, \omega^*)$ via an ANN $\psi = NN(x, m^*, \omega^*; W)$. The corresponding loss function for an observation is the energy expectation value $\langle\psi|H|\psi\rangle/\langle\psi|\psi\rangle$, which is minimized during training. Figure 5(a) shows the exact ground-state wavefunctions for three closely spaced parameter sets used as training observations. Following the same "train-freeze-predict" protocol, the network is trained either on a single observation or on multiple ones via variational learning. Its acquired intuition is then mapped by evaluating prediction accuracy ($R^2 > 0.9$) across a broad range of unseen $(m^*, \omega^*)$. Again, the network learns from single observation, or from multiple observations using the variational learning method. As shown in Figs. 5(b–d), training on a single observation yields a high-accuracy region that is severely localized around the training point. In stark contrast, variational learning on two observations dramatically expands this region, merging the generalization domains of the individual cases. When trained on three observations, the network develops an intuition that generalizes robustly across almost the entire tested parameter space. This progressive expansion of the generalization domain with few instances—consistent with our findings for the hydrogen molecule—directly illustrates how variational learning extracts a unified physical principle from limited data.

**Appendix C: Brachistochrone**

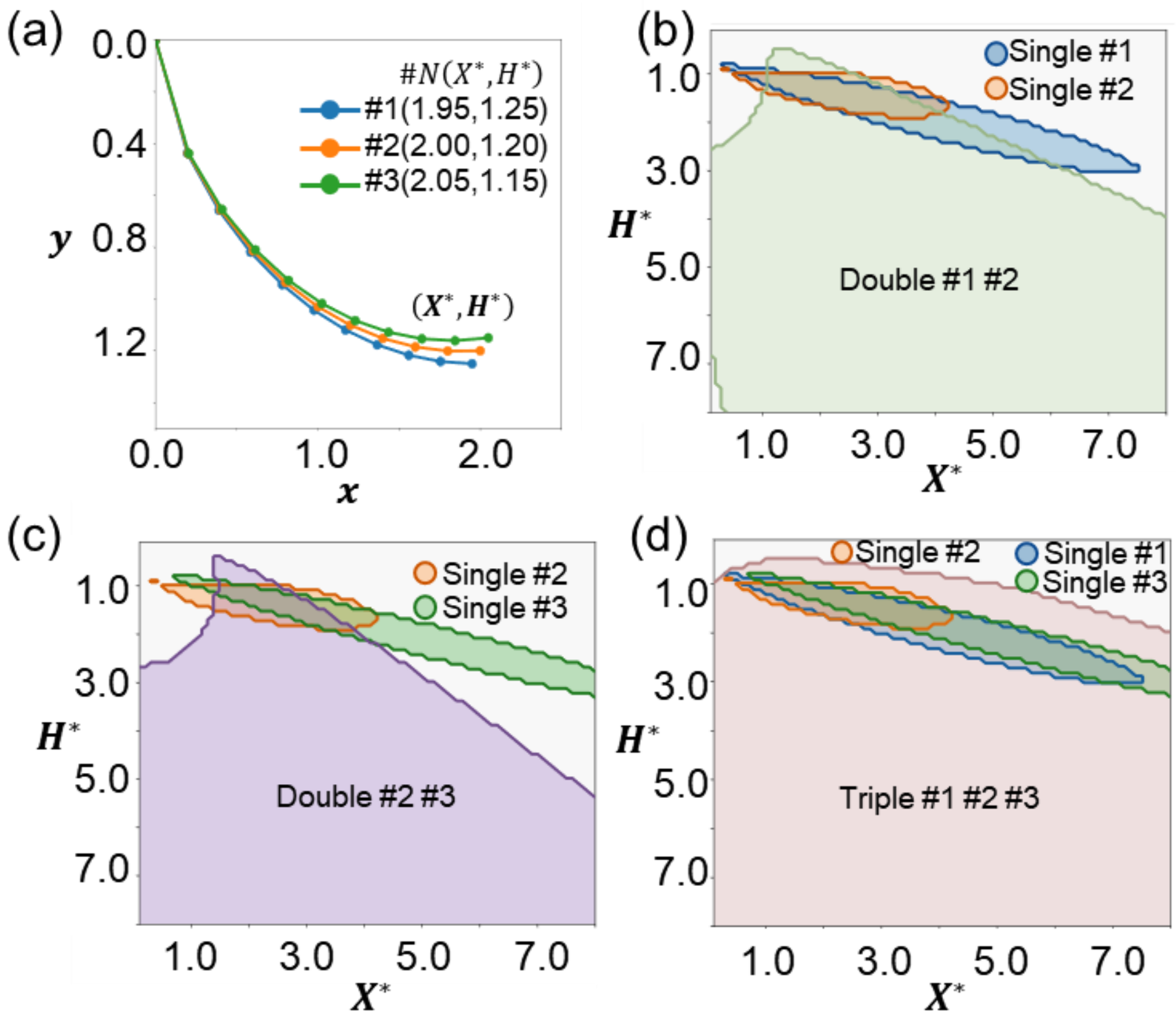

FIG. 6. Accurate physical intuition from a few highly similar observations of brachistochrone curves. (a) Exact brachistochrone trajectories for three terminal points $(X^*, H^*)$ used as training observations (gravity acts downward). (b-d) Generalization performance in the $(X^*, H^*)$ parameter space. Colored regions indicate where the network's prediction accuracy exceeds $R^2$>0.9. Star symbols mark training points. (b) Generalization from single observations (#1, #2) versus variational learning on both (#1, #2). (c) Analogous results for observations #2 and #3. (d) Training on three observations via variational learning enables robust intuition across almost the entire domain, revealing a generalized understanding of the optimal-descent geometry.

We turn to a quintessential classical problem: the brachistochrone curve. Here, the network learns the optimal trajectory $y(x)$ of a particle sliding under gravity from $(0,0)$ to a terminal point $(X^*, H^*)$, via an ANN $y = NN(x, X^*, H^*; W)$. The physics is encoded in a loss function $L(X^*, H^*; W)$ that computes the total travel time, incorporating both the descent integral and a penalty for missing the terminal point. Figure 6(a) shows exact solutions for three closely spaced terminal points, which serve as training observations $(X^*, H^*)$. Results [Figs. 6(b–d)] mirror and extend our quantum findings. Training on a single observation yields a narrow, localized region of competence in parameter space, indicative of overfitting to a single physical instance. Remarkably, variational learning on just two similar observations synthesizes this knowledge, producing a much broader and more connected region of reliable generalization [Figs. 6(b, c)]. When trained on all three observations, the network's intuitive understanding expands to encompass nearly the entire tested domain of terminal points [Fig. 6(d)].

**Appendix D: Projectile motion**

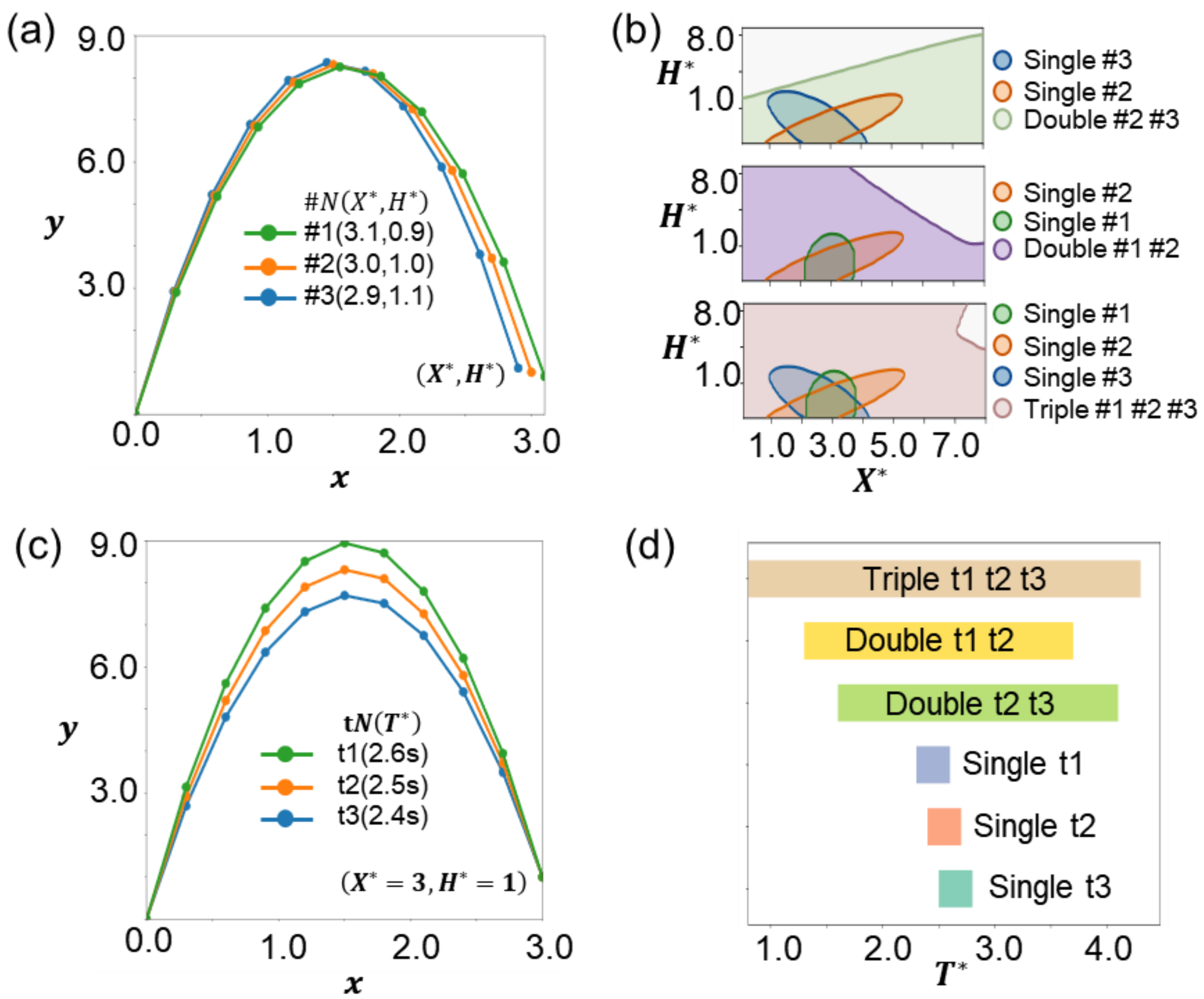

FIG. 7. Variational learning captures both spatial and temporal intuition in projectile motion. (a) Exact trajectories for Task I (fixed time $T^*$, variable target $(X^*, H^*)$) for three highly similar training observations (gravity acts downward). (b) Generalization in Task I: The "good-intuition" region (accuracy $R^2$>0.9) in the target space expands with variational learning from single, double, or triple observations (marked by stars). (c) Exact trajectories for Task II (fixed target $(X^*, H^*)$, variable time $T^*$) for three highly similar training observations. (d) Generalization in Task II: The range of flight time $T^*$ over which accurate predictions are made broadens significantly with variational learning, demonstrating acquired intuition over the temporal parameter.

To complete our demonstration across physics domains, we examine projectile motion, a paradigm of classical mechanics where intuition is often sharpened through practice. Crucially, we design two complementary learning tasks to probe the versatility of variational learning. Task I (Varying Target) concerns learning the trajectory $y(x)$ to hit a variable target $(X^*, H^*)$ within a fixed flight time $T^*$, while Task II (Varying Target) involves learning the trajectory $y(x)$ to hit a fixed target $(X^*, H^*)$ with a variable flight time $T^*$. In both tasks, the trajectory is modeled by double ANNs $[y, x] = [NN_y(t, X^*, H^*, T^*; W), NN_x(t, X^*, H^*, T^*; W)]$, and the physics is encoded in a loss function $L(X^*, H^*, T^*; W)$ based on the Lagrangian action with a boundary penalty. Figures 7(a) and 7 (c) show the ground-truth trajectories for three closely spaced observations in Task I and Task II, respectively. The network is trained, following our established protocol, on either single or multiple observations within each task. The results solidify and extend our framework. For Task I [Fig. 7(b)], variational learning from two or three target points dramatically expands the high-accuracy region in $(X^*, H^*)$ -space, just as in prior systems. More strikingly, for Task II [Fig. 7(d)], where the learned parameter is the scalar time $T^*$, variational learning on multiple time observations substantially broadens the range of over which accurate predictions are made. This shows that the acquired intuition is not merely geometric but encompasses the temporal dimension of the physical law. By successfully handling these two related but distinct tasks, we demonstrate that variational learning extracts a unified, multi-faceted representation of the underlying physics (Newton's laws with gravity), enabling generalization across both spatial and temporal constraints.